\documentclass[12pt]{article}

\usepackage[dvipsnames]{xcolor}
\usepackage{pdfpages}
\usepackage{amsmath}
\usepackage{amsfonts}
\usepackage{upgreek}
\usepackage{sgame}
\usepackage{accents}
\usepackage{amssymb}
\usepackage{tikz-cd}
\usepackage{float}
\usepackage[numbers]{natbib}   
\usepackage{dutchcal}
\usepackage{mathtools}
\usepackage{amsthm}
\usepackage{mdframed}
\usepackage{amssymb}
\usepackage{enumitem}
\usepackage{epigraph}
\usepackage{sectsty}

\newtheorem{theorem}{Theorem}[section]
\newtheorem{proposition}[theorem]{Proposition}
\newtheorem{lemma}[theorem]{Lemma}
\newtheorem{corollary}[theorem]{Corollary}
\newtheorem{claim}[theorem]{Claim}

\theoremstyle{definition}

\newtheorem{definition}[theorem]{Definition}
\newtheorem{example}[theorem]{Example}

\newtheorem{question}[theorem]{Question}

\newtheorem{remark}[theorem]{Remark}

\newtheorem{aside}[theorem]{Aside}
\newtheorem{note}[theorem]{Note}

\usepackage{hyperref}
\hypersetup{
    colorlinks=true,
    linkcolor=Maroon,
    filecolor=magenta,
    urlcolor=Mulberry,
    citecolor=MidnightBlue,
}


\begin{document}
\author{Isaac M. Sonin\thanks{Department of Mathematics, University of North Carolina at Charlotte} \and Mark Whitmeyer\thanks{Corresponding Author; Department of Economics, University of Texas at Austin \newline Email: \href{mailto:mark.whitmeyer@utexas.edu}{mark.whitmeyer@utexas.edu}.} }

\title{Some Nontrivial Properties of a Formula for Compound Interest}

\maketitle

\begin{abstract}
We analyze the classical model of compound interest with a constant per-period payment and interest rate. We examine the outstanding balance function as well as the periodic payment function and show that the outstanding balance function is not generally concave in the interest rate, but instead may be initially convex on its domain and then concave.
\end{abstract}

\newpage

\section{Introduction}
\setlength{\epigraphwidth}{5in} 
\begin{epigraphs}
\qitem{When men's minds began to be more enlarged, when true religion and real liberty revived, commerce grew again into credit; and again introduced with itself its inseparable companion, the doctrine of loans upon interest.}%
{---\textsc{Commentaries on the Laws of England  (William Blackstone)}}
\end{epigraphs}

The formula for compound interest has been known to humankind for thousands of years, dating back even to ancient Sumer, where over 4000 years ago merchants were lending and calculating compound interest on loans (Muroi, 2015) \cite{sumer}. Unsurprisingly, the enmity and resentment provoked by compound interest is equally ancient. Lending at interest has been regulated by governments and traditions in many countries, and has naturally drawn ire from many religions as well as historical figures such as Aristotle, Plato,  William Shakespeare, and Karl Marx, to name just a few. On the other hand, there are stronger arguments that compound interest should be considered one of the engines driving the wheels of civilization itself.\footnote{For more details about the importance of interest throughout history, see the excellent article by Hudson (2000) \cite{hud}, and books by Homer and Sylla (1963) \cite{homer}, and Goetzmann (2016) \cite{money}.}

Despite the significance of compound interest--and the length of time it has resided in humanity's sphere of knowledge--some of its fundamental traits have escaped characterization. Hence, the purpose of this note is to discuss properties of two important functions related to compound interest. We look at simple loan contracts in which the amount borrowed, $C$, must be repaid over $T$ periods with
the deterministic and constant interest rate $r$ and a constant periodic payment $P(r,T)$. Then

\[\tag{$1$}\label{1}
C= \frac P {1+r}+\frac P {(1+r)^2} + \cdots +\frac P {(1+r)^T} = P \bigg[\frac{(1+r)^{T}-1}{r(1+r)^{T}}\bigg]  \]
where $P = P(r,T)$ is the {\it Periodic Payment}.

The {\it Outstanding Balance}, $B=B(r,T,t)$, is the amount due at moment $t$, $0 \leq t \leq T$ if periodic payment is stopped at moment $t$. That is, $B$ is defined by the equality

\[\label{3}\tag{$2$}1= \frac P {1+r}+\frac P {(1+r)^2} +\cdots + \frac P {(1+r)^t}+\frac B {(1+r)^t}\]

In our first result, Proposition \ref{prop21}, we show that for a fixed $T$ and $t$, the outstanding balance function is an increasing function of $r$. If $T - 2t + 3 \geq 0$ then $B$ is a concave function of $r$, and it is convex-concave otherwise. Next, we look at the periodic payment, $P(r,T)$, as a function of the interest rate, $r$, and the amortization interval, $T$. In Proposition \ref{prop22}, we establish that for a fixed $T$, the periodic payment is increasing and convex in $r$.

Results concerning these functions were used in Dunn and Spatt (1999) \cite{spatt}. In Lemma 2 (Proposition \ref{prop22} in this paper), they characterize the periodic payment function but omit the proof. In Lemma 3, the authors write that the outstanding balance function is a strictly concave function of the interest rate. However, here we show that this is true only for certain values of $T$ and $t$, contrary to the assertion that it is true for all values.

Finally, we would like to mention that the concept of interest rate is deeply connected with the important and not completely understood concept of Internal Rate of Return. See the classical papers, Arrow and Lehtani (1969) \cite{arrow}, Dorfman (1981) \cite{dor}, and Samuelson (1937) \cite{sam}; as well as more recent work, Atsumi (1991) \cite{at}, Hazen (2003) \cite{haze}, and Sonin (1995) \cite{sonin}.

\section{Results}
 We assume from now on that $C=1$ and rearrange Expression \ref{1} to obtain
\[\tag{$3$}\label{2}
P(r,T) = \frac{r(1+r)^T}{(1+r)^T-1} =r+\frac r{(1+r)^T-1}
\]
We may rearrange the terms in the outstanding balance function, $B(r,T,t)$ to obtain

\[\label{4}\tag{$4$}B(r,T,t)= \frac {(1+r)^T -(1+r)^t} {(1+r)^T -1} = 1- \frac {(1+r)^t-1}
{(1+r)^T-1}\]

We call a function {\it convex-concave} if on the interval of its definition it is first convex on some nonempty interval and after that concave.

\begin{proposition}\label{prop21}
The outstanding balance function, $B(r,T,t)$, is an increasing function of an interest rate $r$. Moreover, it is a concave function of $r$ if $T-2t+3\geq 0$, and a convex-concave function of $r$ otherwise.
\end{proposition}

\begin{proposition}\label{prop22}
The periodic payment function, $P(r,T)$, is an increasing convex function of interest rate $r$ for any given fixed $T$.
\end{proposition}

The proofs are straightforward but tedious exercises in calculus. We will describe the main steps a bit later (Section \ref{proofs}), but first briefly mention some possible applications of these facts.

In economics and finance the convexity properties of functions matter mainly when stochasticity is involved. If, in a simplified example, a borrower faces a prospect of having a loan with an interest rate equal to $r \pm a$ with equal probabilities, then Proposition \ref{prop22} implies immediately that an {\it Expected} periodic payment will be higher than the periodic payment for an interest rate $r$. If, for the same example, the periodic payments are stopped at moment $t,$ then whether the expected outstanding balance will be lower or higher depends on the sign of the expression $T-2t+3$. 

Dunn and Spatt (1999) \cite{spatt} use the (purported) concavity of the outstanding balance function in order to establish that for three callable loans with the same maturity, the (discount) points are a strictly concave function of the interest rate $r$ (Proposition 5). One particular strength of the Dunn and Spatt paper is the parsimonious nature of the model, which allows for the authors to derive clear economic predictions absent particular parametric assumptions. More sophisticated related papers include Kau et al. (1992) \cite{kau}, which seeks to provide a valuation model for fixed rate mortgages where the change in valuation of the house and the interest rate follow continuous time stochastic processes; and Mele (2003) \cite{mele}, which characterizes some simple relationships between bond prices and the short term interest rate. As in \cite{kau}, the interest rate (as well as the short-term rate's volatility) are continuous time Markov processes, and the author provides conditions under which bond prices are strictly decreasing and convex in the (short-term) interest rate.

One final paper that bears mentioning is Berk (1999) \cite{berk}, in which the author derives a modified net present value investment rule, that can be used in situations where an investor has an option to delay his investment decision.


\subsection{Proof of Propositions \ref{prop21} and \ref{prop22}}\label{proofs}

For convenience, we use the notations $1+r=x$, $T=n$, and $t=k$. Then, expressions (\ref{4}) and (\ref{2}) can be rewritten as

\[\label{5} \tag{$5$} B(x,n,k)=\frac {x^n-x^k} {x^n-1} = 1 - \frac {x^k -1} {x^n-1}\]

and

\[\label{6} \tag{$6$} P(x,n)= \frac {x^{n+1}-x^n} {x^n-1}= x - 1 + \frac {x -1} {x^n-1} =x - B(x,n,1)\]

It is easy to check that $B(1,n,k)=(n-k)/n$ and $P(1,n)=1/n$. To prove Proposition \ref{prop21} we have to analyze the sign of the first and second derivatives of $B(x,n,k)$, which are $B_{x}$, and $B_{xx}$, respectively. We have

\[\begin{split}
    B_{x}(x,n,k) = - \frac{x^{k-1}\big(x^{n}(k-n) + nx^{n-k} - k\big)}{(x^{n}-1)^{2}} \equiv - \frac{x^{k-1}}{(x^{n}-1)^{2}}B_{1}(x)
\end{split}\]

Clearly, $B_{1}(1)=0,$ and $B_{1}^{'}(x)=x^{n-k-1}n(k-n)(x^k-1)<0 $ for all $x>1$. Hence, since $B_{1}(x)$ is strictly negative, $B_{x}$ is strictly positive, i.e. $B(x, n, k)$ is an increasing function of $x$. Next, we take the second derivative of $B$ with respect to $x$, $B_{xx}$:

\[\label{7} \tag{$7$}
    B_{xx}(x,n,k) = - \frac{x^{k-2}\big(Ax^{2n} + Cx^{2n-k} + Dx^{n} + Ex^{n-k} + F\big)}{(x^{n}-1)^{3}} \equiv - \frac{x^{k-2}}{(x^{n}-1)^{3}}B_{2}(x)\]
where $A = (n-k)(n-k+1)$, $C = -n(n+1)$, $D = -2k(k-1)+2kn+n(n-1)$, $E = -n(n-1)$, and $F = k(k-1)$. We introduce the following sequence of functions, $B_{3}(x)$, $B_{4}(x)$, and $B_{5}(x)$, where $x^{n-k-1}B_{3}(x) = B_{2}^{'}(x)$, $nx^{k-1}B_{4}(x) = B_{3}^{'}(x)$, and $x^{n-k-1}B_{5}(x) =  B_{4}^{'}$. The full details are left to Appendix \ref{app}.

Direct calculations show that $B_{2}(1) = B_{3}(1) = B_{4}(1)=0$, and that $B_{5}(1)= kn(n-k)(n-2k+3)$. Moreover, $B_{5}(x)=2n(n+k)Ax^k+(n-k)(2n-k)C$. Accordingly, since $A>0$, function $B_{5}(x)$ is increasing in $x$, positive for large $x$, and approaches $+\infty$ as $x$ becomes infinitely large.

If $n-2k+3\geq 0$ then $B_{5}(x)$ is positive for all $x > 1$, and since $B_{2}(1)=B_{3}(1)=B_{4}(1)=0$ the same is true for functions $B_{2}$, $B_{3}$, and $B_{4}$. Thus by (\ref{7}) the second derivative of $B(x)$ is negative for all $x>1$, i.e. $B(x)$ is a concave function for all $x\geq 1$.

On the other hand, if $n-2k+3<0$ then function $B_{5}(x)(x)$ is negative on some nonempty interval $[1,r_{0})$, positive after $r_{0}$, and approaches $+\infty$ in the limit. Therefore the same is true for functions $B_{2}$, $B_{3}$, and $B_{4}$ for some nonempty intervals, and it means that the second derivative of function $B(x)$ is positive on some interval and after that negative. Hence, for these values of $n$ and $k$, function $B(x)$ is convex-concave. This proves Proposition \ref{prop21}. $\qedsymbol$

The inflexion point where the concavity replaces the convexity can be found numerically as a solution of a polynomial equation $B_2(x)=0$ (see Expression \ref{7}).

Next, to prove Proposition \ref{prop22}, we take the partial derivative of $P(x,n)$ with respect to $x$, $P_{x}$:

\[\begin{split}
    P_{x}(x,n) = \frac{x^{n-1}\big(x^{n+1} - (n+1)x + n\big)}{(x^{n}-1)^{2}} \equiv \frac{x^{n-1}}{(x^{n}-1)^{2}}P_{1}(x)
\end{split}\]

Clearly, $P_{1}(1)=0$ and $P_{1}^{'}(x)=(n+1)(x^n-1)>0$ for all $x>1$ and hence $P_{x}>0$ i.e. $P(x,n)$ is increasing in $x$. Finally, from Proposition \ref{prop21}, $B(x,n,1)$ is concave (since $T-2+3>0$). Then, $-B(x,n,1)$ is convex and by Expression \ref{6} function $P(x,n)$ is also
convex. $\qedsymbol$

\bibliography{sample.bib}

\appendix

\section{Proposition \ref{prop21} Calculations}\label{app}

Differentiating function $B_{2}(x)$ we obtain

\[\begin{split}
    B_{2}^{'}(x) &= 2nAx^{2n-1} + (2n-k)Cx^{2n-k-1} + nDx^{n-1} + (n-k)Ex^{n-k-1}\\
    &= x^{n-k-1}\big(2nAx^{n+k} + (2n-k)Cx^{n} + nDx^{k} + (n-k)E\big) \equiv x^{n-k-1}B_{3}(x)
\end{split}\]
Then

\[\begin{split}
    B_{3}^{'}(x) &= 2n(n+k)Ax^{n+k-1} + (2n-k)nCx^{n-1} + knDx^{k-1}\\
    &= nx^{k-1}\big(2(n+k)Ax^{n} + (2n-k)Cx^{n-k} + kD\big) \equiv nx^{k-1}B_{4}(x)
\end{split}\]
and

\[\begin{split}
    B_{4}^{'}(x) &= 2n(n+k)Ax^{n-1} + (2n-k)(n-k)Cx^{n-k-1}\\
    &= x^{n-k-1}\big(2n(n+k)Ax^{k} + (2n-k)(n-k)C\big) \equiv x^{n-k-1}B_{5}(x)
\end{split}\]

\end{document}